\documentclass{jpsj-suppl}
\usepackage{txfonts} 

\title{Beta-decay Studies in $N\approx Z$ Nuclei Using
No-Core Configuration-Interaction Model}

\author{Wojciech \textsc{Satu{\l}a}$^{1}$, Jacek \textsc{Dobaczewski}$^{1,2}$, and Maciej \textsc{Konieczka}$^{1}$}

\inst{$^{1}$Institute of Theoretical Physics, Faculty of Physics, University of Warsaw, ul. Ho\.za 69, PL-00-681 Warsaw, Poland \\
$^{2}$Department of Physics, P.O. Box 35 (YFL), University of Jyv\"askyl\"a, FI-40014  Jyv\"askyl\"a, Finland}

\email{satula@fuw.edu.pl}

\recdate{June 23, 2011}

\abst{The no-core configuration-interaction model based on the
isospin- and angular-momentum projected density functional formalism
is introduced. Two applications of the model are presented: (i)
determination of spectra of $0^+$ states in $^{62}$Zn and (ii)
determination of isospin-symmetry-breaking corrections to
superallowed $\beta$-decay between isobaric-analogue $0^+$ states in
$^{38}$Ca and $^{38}$K. It is shown that, without readjusting a
single parameter of the underlying Skyrme interaction, in all three
nuclei, the model reproduces the $0^+$ spectra surprisingly well.}

\kword{density functional theory, configuration-interaction model, superallowed Fermi $\beta$-decay, isospin symmetry breaking}

\newcommand{\thalf}{\tfrac{1}{2}}
\newcommand{\ttalf}{\tfrac{3}{2}}

\begin{document}
\maketitle

\section{Introduction}

Superallowed Fermi $\beta$-decays between the isobaric analogue states,
[$I=0^+,T=1]\rightarrow [I=0^+,T=1]$, provide the most precise values
of the vector coupling constant $G_V$ and leading element $V_{ud}$ of
the Cabibbo-Kobayashi-Maskawa (CKM) flavour-mixing matrix, which are
critical for stringent tests of weak-interaction flavor-mixing sector
of the Standard Model of particle physics. In particular, these data
are needed for testing unitarity of the CKM matrix, violation of
which may signalize {\it new physics\/} beyond the Standard Model,
see~\cite{(Tow10a)} and refs. quoted therein.

In testing the Standard Model, precision is of utmost importance. Only
the $\beta$-decays, for which the reduced life-times, $ft$, are measured
with accuracy better than half a percent, can be used for that
purpose. At present, thirteen such cases are known in nuclei, ranging
in mass from $A$=10 to $A$=74. The extraction of $G_V$ and $V_{ud}$
is not solely dependent on experimental data but also requires
theoretical input in the form of radiative and many-body
corrections to the experimental $ft$ values. The corrections are
small, of the order of a percent, but are critical for the applicability of
the entire method, because it relies on the so-called conserved vector
current hypothesis (CVC).  The CVC hypothesis assumes independence of
the vector current on nuclear medium, and must be verified {\it a
priori\/} by investigating mass independence of the corrected reduced
life-times defined as:
\begin{equation}\label{eq:Ft}
   {\cal F}t \equiv ft(1+\delta_{\rm R}^\prime)(1+\delta_{\rm NS} -\delta_{\rm C})
  = \frac{K}{2 G_{\rm V}^2 (1 + \Delta^{\rm V}_{\rm R})} \approx {\rm const.}\, ,
\end{equation}
where $K/(\hbar c)^6 = 2\pi^3 \hbar \ln 2 /(m_{\rm e} c^2)^5 =
8120.2787(11)\times 10^{-10}$\,GeV$^{-4}$s  is a universal constant.
Symbols $\delta_{\rm R}^\prime, \delta_{\rm NS}, \Delta^{\rm V}_{\rm
R}$ are the radiative corrections while $\delta_{\rm C}$ stands for
the isospin symmetry-breaking (ISB) correction to the Fermi matrix
element:
\begin{equation}\label{eq:ME}
|M_{\rm F}^{(\pm )}|^2 = |\langle I=0^+, T\approx 1, T_z = \pm 1 | \hat T_\pm | I=0^+, T\approx 1, T_z = 0\rangle |^2 = 2 (1- \delta_{\rm C} ).
\end{equation}

Since the isospin symmetry is weakly broken, mostly by the Coulomb
interaction that polarizes the entire nucleus, microscopic
calculation of the ISB corrections is a challenging task. Capturing
a delicate equilibrium between the hadronic and Coulomb effects is fully
possible only within {\it no core\/} approaches. This, in
heavier nuclei, reduces the possible choices to formalisms rooted in the
density functional theory (DFT). However, as it was recognized already in the
70's~\cite{(Eng70)}, to determine the magnitude
of isospin impurities, the self-consistent mean-field (MF) approaches
cannot be directly applied, because of a spurious mixing caused by the
spontaneous symmetry-breaking effects. This observation hindered
theory from progress in the field for decades.

To overcome these problems, over the last few years we have developed
a no-core multi-reference DFT, which involves the isospin-
and angular-momentum projections of Slater determinants representing
the $0^+$ triplet states in mother and daughter
nuclei~\cite{(Sat11),(Sat12)}. The formalism, dubbed static, was
specifically designed to treat rigorously the conserved rotational
symmetry and, at the same time, tackle the explicit breaking of the
isospin symmetry due to the Coulomb field. Recently, by allowing for
mixing of states that are projected from self-consistent Slater
determinants representing low-lying (multi)particle-(multi)hole
excitations, we have extended the model to the so-called dynamic
variant~\cite{(Sat14)}. The model belongs to the class of the {\it no core\/}
configuration-interaction approaches, with the two-body short-range
(hadronic) and long-range (Coulomb) interactions treated on the same
footing. It is based on a truncation scheme dictated by the
self-consistent deformed Hartree-Fock (HF) solutions. The model can
be used to calculate spectra, transitions, and $\beta$-decay matrix
elements in any nuclei, irrespective of their mass and neutron- and
proton-number parities.

The aim of this work is to present this novel theoretical framework
along with preliminary results for the low-spin spectra and
$\beta$-decay matrix elements in selected $N\approx Z$ nuclei. The
first applications of the model to the low-lying spectra in $^{32}$Cl
and $^{32}$S have been published in~\cite{(Sat14)}.
Hereafter, we focus on nuclei relevant to high-precision tests of the
weak-interaction flavor-mixing sector of the Standard Model. In this perspective,
we discuss the spectrum of $0^+$ states in $^{62}$Zn, which was
reassigned in a recent experiment~\cite{(Lea13)}, and is now posing a
challenge to theory. We also briefly overview preliminary attempts
and difficulties arising in determining the ISB correction to the
superallowed $^{62}$Ga$\rightarrow ^{62}$Zn $0^+\rightarrow 0^+$
$\beta$-decay, which is strongly model dependent. We also
present preliminary results for the ISB correction to the Fermi
matrix element corresponding to the $^{38}$Ca$\rightarrow ^{38}$K
transition. In our static calculations, the case of $A$=38 was
excluded from the canonical pool of superallowed data. This was
because of the anomalously large ISB correction, caused by
uncontrolled mixing of the $2s_{1/2}$ and $1d_{3/2}$ orbits, which for
the SV$_{\rm T}$ Skyrme {\it true\/} interaction are almost
degenerate. The SV$_{\rm T}$ interaction is the SV
functional~\cite{(Bei75)} augmented with the tensor terms, see
discussion in~\cite{(Sat14a)}.

The paper is organized as follows. In Sec.~\ref{sec2}, the basics of
our dynamical model are briefly sketched. In Sec.~\ref{sec3},
preliminary numerical results concerning spectrum of $0^+$ states in
$^{62}$Zn and the ISB corrections for $^{38}$Ca$\rightarrow$$^{38}$K
Fermi transitions are presented. The paper is summarized in
Sec.~\ref{sec4}.

\section{No-core configuration-interaction model}\label{sec2}

The static variant of our model is based on the double projection, on
isospin and angular momentum, of a single Slater determinant. In an
even-even nucleus, the Slater determinant representing the
ground-state is uniquely defined. In an odd-odd nucleus, the
conventional MF theory that gives Slater determinants separably for
neutrons and protons faces problems. First, there is no single Slater
determinant representing the $I=0^+, T=1$ state,
see~\cite{(Sat12),(Sato13)}. In our approach, this obstacle is
removed by projecting from the so-called anti-aligned Slater
determinant. This configuration, by construction, has no net
alignment and manifestly breaks the isospin symmetry, being an almost
fifty-fifty mixture of the $T=0$ and $T=1$ states. In this way, the
needed $T=1$ component can be recovered. The problem is, however,
that the anti-aligned states are not uniquely defined. In the general
case of a triaxial nucleus, there exist three linearly-dependent
Slater determinants, built of valence neutron and proton
single-particle states that carry angular momenta aligned along the X
($|\varphi^{\rm (X)}\rangle$), Y ($|\varphi^{\rm (Y)}\rangle$), or Z
($|\varphi^{\rm (Z)}\rangle$) axes of the intrinsic frame of
reference or, respectively, along the long, intermediate, and short
axes of the core. In our calculations, no tilted-axis anti-aligned
solution was found so far.

In the static approach, the only way to cope with this ambiguity is
to calculate three independent $\beta$-decay matrix elements and to
take the average of the resulting $\delta_{\rm C}$ values. Such a
solution is not only somewhat artificial, but also increases the
theoretical uncertainty of the calculated ISB corrections. This
deficiency motivated our development of the dynamic model, which
allowed for mixing states projected from the three reference states
$| \varphi^{\rm (k)}\rangle$ for k=X, Y, and Z, with the mixing matrix
elements derived from the same Hamiltonian that was used to calculate
them. The dynamic model further evolved towards a full
no-core configuration-interaction (NCCI) model, in which we allow for
mixing states projected from different low-lying
(multi)particle-(multi)hole Slater determinants $|\varphi_i\rangle$.
This final variant has all features of the {\it no core\/} shell
model, with two-body effective interaction (including the Coulomb
force) and a basis-truncation scheme dictated by the self-consistent
deformed Hartree-Fock solutions.

The computational scheme proceeds in four
major steps:
\begin{itemize}

\item
First, a set of relevant low-lying (multi)particle-(multi)hole HF
states $\{ \varphi_i \}$ is calculated along with their HF energies
$e^{{\rm (HF)}}_i$,  which form a subspace of reference states for
subsequent projections.

\item
Second, the  projection techniques are applied to the set of states
$\{ \varphi_i \}$, so as to determine the family of states $\{
\Psi_{TIK}^{(i)}\}$ having good isospin $T$, angular momentum $I$, and
angular-momentum projection on the intrinsic axis $K$.

\item
Third, states $\{\Psi_{TIK}^{(i)}\}$ are mixed, so as to properly take into
account the $K$ mixing and Coulomb isospin mixing -- this gives the set of
good angular-momentum states $\{\Psi_I^{(i)}\}$ of the static model~\cite{(Sat11),(Sat12)}.

\item
Finally, the mixing of, in general, non-orthogonal states $\{
\Psi_I^{(i)}\}$ for all configurations $i$ is performed by solving the Hill-Wheeler
equation in the collective space spanned  by the natural states
corresponding to non-zero eigenvalues of their norm matrix, that is, by
applying the same technique, which was used to handle the
$K-$mixing alone~\cite{(Dob09ds)}.

\end{itemize}

The numerical stability of the method is affected by necessary
truncations of the model space, namely, numerically unstable
solutions are removed by truncating either the high-energy states
$\{\Psi_I^{(i)} \}$ or the {\it natural states\/} corresponding to
small eigenvalues of the norm matrix, or by applying both truncations
simultaneously. Although such truncation procedure gives reliable
values of the energy, the results shown below must still be
considered as preliminary.

\section{Numerical results}\label{sec3}

The CKM matrix element $|V_{ud}|=0.97397(27)$ obtained with a set of
the ISB corrections calculated using the double-projected DFT method~\cite{(Sat12)}
agrees very well with the result obtained by Towner and Hardy
(TH)~\cite{(Tow08)}, $|V_{ud}|=0.97418(26)$, obtained within methodology based
on the nuclear shell-model combined with Woods-Saxon mean-field
(SM+WS) wave functions.  Both values result in  the unitarity of the
CKM matrix up to 0.1\%. It is gratifying to see that also individual
DFT values of $\delta_{\rm C}$ are consistent within 2$\sigma$ with
the values calculated in Ref.~\cite{(Tow08)} (see Fig.~7 of
Ref.~\cite{(Sat12)}). This holds up to three exceptions of the ISB
corrections to $^{10}$C$\rightarrow$$^{10}$B,
$^{38}$K$\rightarrow$$^{38}$Ar, and $^{62}$Ga$\rightarrow$$^{62}$Zn
transitions. The two latter mass numbers, more precisely transitions
$^{38}$Ca$\rightarrow$$^{38}$K and $^{62}$Ga$\rightarrow$$^{62}$Zn,
are analyzed below using the newly developed NCCI approach.  It is worth mentioning here
that the mutually consistent DFT and TH results are at variance with the
RPA-based study of Ref.~\cite{(Lia09)}, which gives
systematically smaller values of $\delta_{\rm C}$ and, in turn,
considerably smaller value of matrix element $V_{ud}$.

\subsection{No-core configuration-interaction calculations for $0^+$ states in $^{62}$Zn}

\begin{figure}[tbh]
\centering\includegraphics[width=0.5\textwidth]{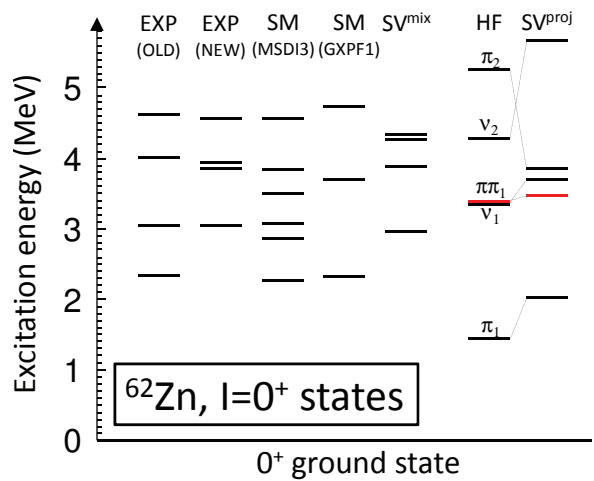}
\caption{The first five columns show low-energy $I$=0$^+$ states in
$^{62}$Zn listed in Table~\protect\ref{t1}. Columns marked HF and
SV$^{{\rm proj}}$ show MF results obtained for the six HF
configurations and those for the $0^+$ states projected from the HF
configurations before the mixing, respectively.}
\label{f1}
\end{figure}

A large difference between the ISB corrections to the
$^{62}$Ga$\rightarrow$$^{62}$Zn Fermi matrix element, calculated using
the DFT and SM+WS approaches, is one of the motivations to
undertake the NCCI studies of the participating
nuclei. Interestingly, nucleus $^{62}$Zn has been recently remeasured
by the TRIUMPH group~\cite{(Lea13)}, and its spectrum of
low-lying $0^+$ states is now posing a great challenge to theory, as
shown in the Tab.~\ref{t1} and Fig.~\ref{f1}. Both the table and
figure also include  results of our NCCI study, which
involves the mixing of $0^+$ states projected from six reference states.
They comprise the deformed ground state (g.s.) and
five low-lying excited HF configurations, including two lowest proton
($\pi_1$ and $\pi_2$) and two lowest neutron ($\nu_1$ and $\nu_2$)
p-h excitations, and the lowest proton-proton 2p-2h configuration, all
calculated with respect to the g.s.

\begin{table}[tbh]
\caption{Excitation energies of $0^+$ states in $^{62}$Zn up to
5\,MeV. First two columns show old and new experimental data,
see~\cite{(Lea13)} for details. Next three columns collect the
results of shell-model calculations using MSDI3~\cite{(Koo77)},
GXPF1~\cite{(Hon02)}, and GXPF1A~\cite{(Hon04)} interactions,
respectively. Last column shows the results obtained in this work
using six reference Slater determinants described in the text.
All values are in keV.}
\label{t1}
\centering\begin{tabular}{rrrrrr}
\hline
OLD         &  NEW        &  MSDI3 &   GXPF1 &  GXPF1A  &  SV$^{{\rm mix}}$ \\
\hline
2341.95(2)  &             &   2263 &    2320 &    2094  &                   \\
            &             &   2874 &         &    2811  &                   \\
3042.9(8)   &  3045.5(4)  &   3071 &         &    3457  &    2953           \\
            &  3862(2)    &   3513 &    3706 &    3682  &    3884           \\
4008.4(7)   &  3936(6)    &   3833 &         &    3991  &    4263           \\
            &             &        &         &    4444  &                   \\
4620(20)    &  4552(9)    &   4551 &    4729 &    4643  &    4347           \\
\hline
\end{tabular}
\end{table}

\begin{figure}
\begin{center}
\begin{minipage}[t]{0.48\textwidth}
\begin{center}
\includegraphics[width=0.98\textwidth]{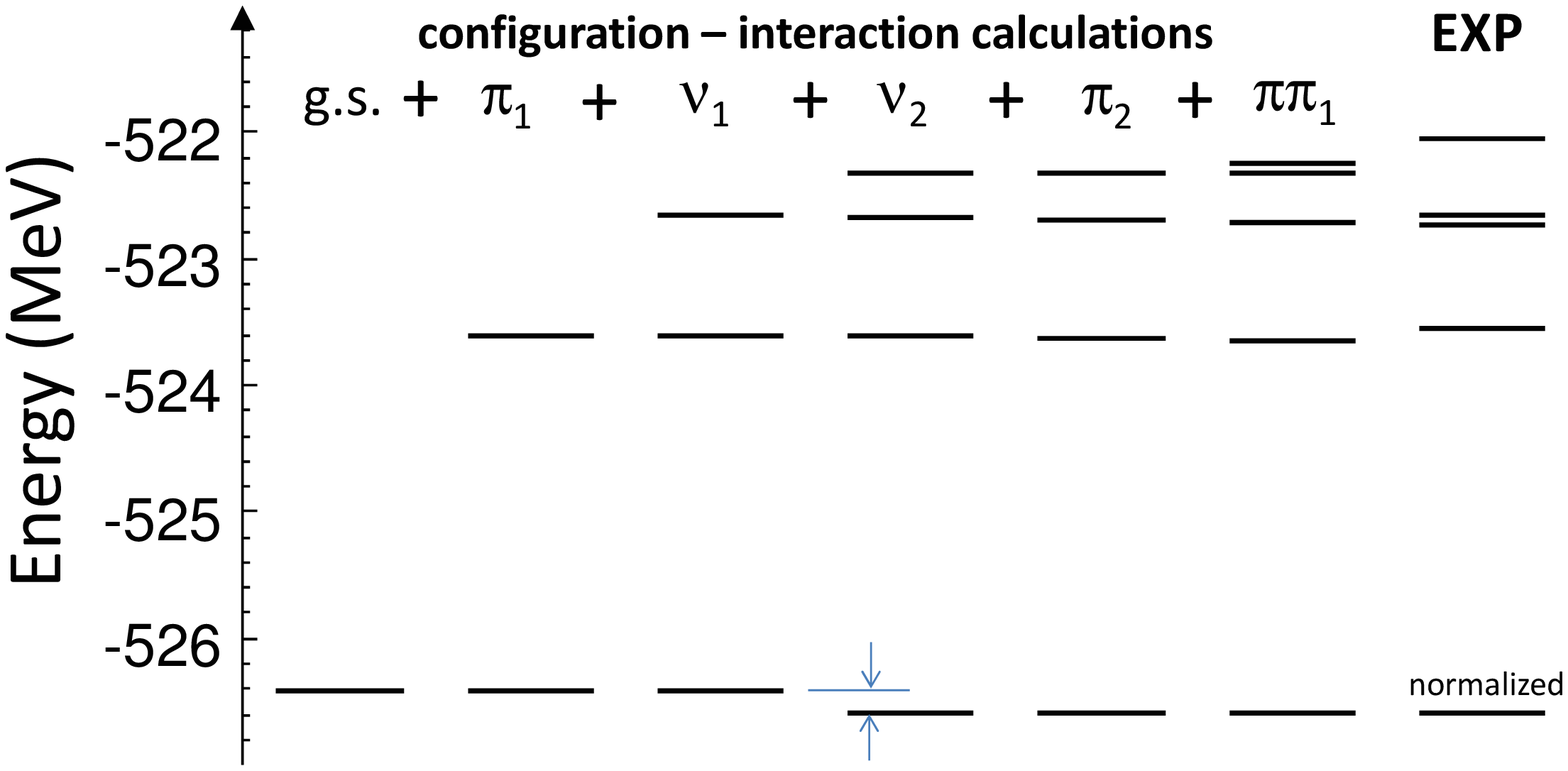}
\caption{\label{f2}Stability of the absolute energies of the calculated $0^+$
states in $^{62}$Zn with respect to number of reference
configurations included in the calculations.}
\end{center}
\end{minipage}\hspace{0.03\textwidth}%
\begin{minipage}[t]{0.48\textwidth}
\begin{center}
\includegraphics[width=0.58\textwidth]{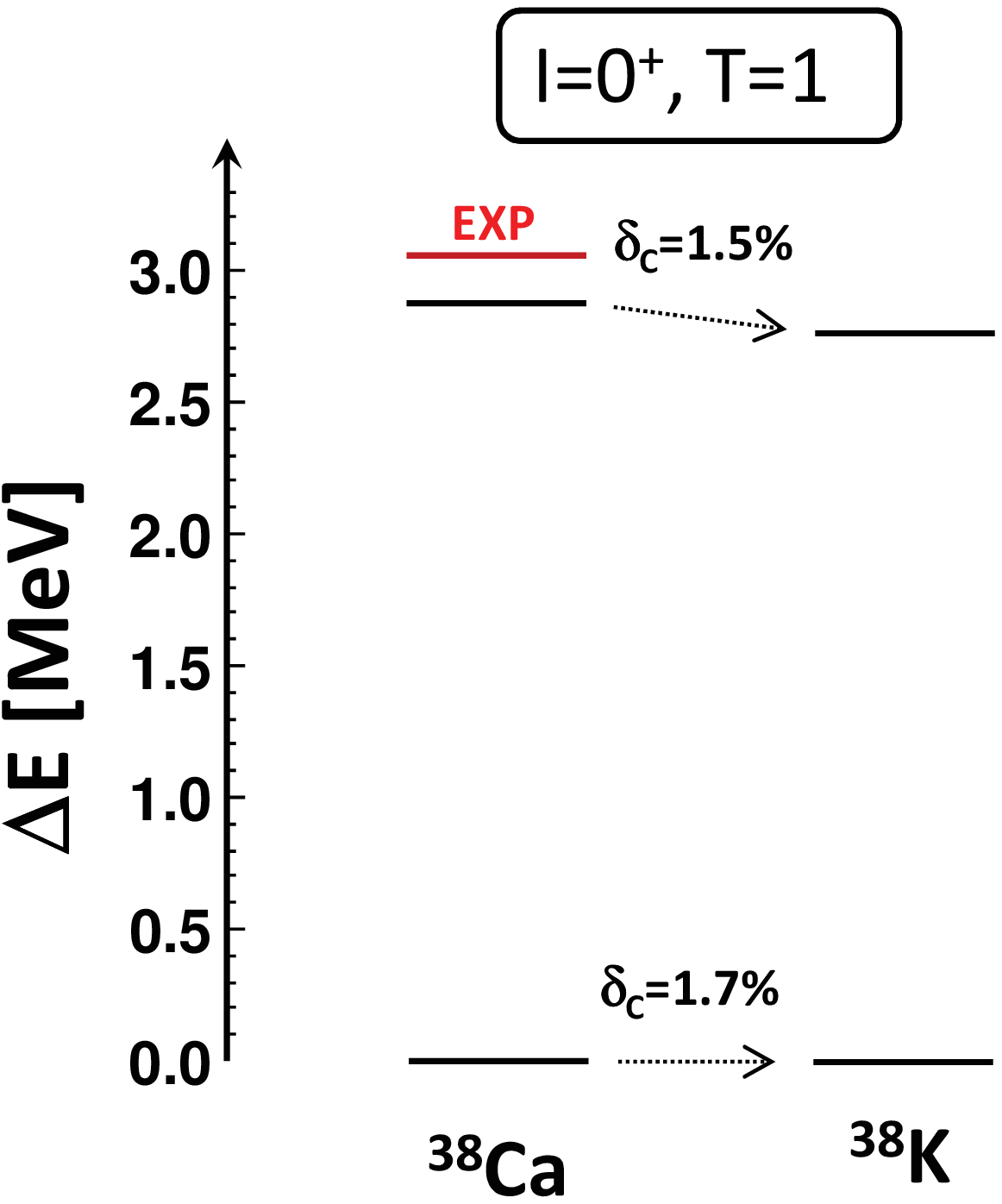}
\caption{\label{f3}Calculated energies of the
$0_2^+$ relative to $0_1^+$ states in $^{38}$Ca
and  $^{38}$K, and the ISB corrections to the corresponding $0_1^+\rightarrow
0_1^+$ and $0_2^+\rightarrow 0_2^+$ transitions. Empirical excitation
energy of the $0_2^+$ state in $^{38}$Ca is also shown.}
\end{center}
\end{minipage}
\end{center}
\end{figure}

It is gratifying to observe that our model is able to capture,
without adjusting a single parameter, the spectrum of $0^+$ states in
$^{62}$Zn very accurately, even better than the state-of-the-art SM
calculations. Moreover,  as shown in Fig.~\ref{f2}, the calculated
spectrum of the $0^+$ states in $^{62}$Zn is relatively stable with
respect to increasing the number of reference configurations.
The last two columns of Fig.~\ref{f1} illustrate the importance
of symmetry restoration and configurations mixing.

Unfortunately, the calculated corrections $\delta_{\rm C}$ are
sensitive to tiny admixtures to the wave function and, and present,
the calculated values are not stable. For example, by adding the
$0^+$ state projected from configuration $\nu_2$, one changes the
absolute g.s.\ energy of $^{62}$Zn by only $\approx$200keV, but at
the same time $\delta_{\rm C}$ changes by $\approx$4\%. Such a large
change of $\delta_{\rm C}$ is probably entirely artificial,
reflecting the fact that the spaces of states used to calculate the
parent and daughter nuclei do not match.

\subsubsection{ISB correction to $^{38}$K$\rightarrow$$^{38}$Ca transition}

In the static DFT calculations, the ISB correction to the
$^{38}$K$\rightarrow$$^{38}$Ar and $^{38}$Ca$\rightarrow$$^{38}$K
superallowed transitions turned out to be unphysically
large~\cite{(Sat11)}, and were disregarded. The reason could be traced
back to unphysical values of the single-particle (s.p.) energies of the  $2s_{1/2}$ and $1d_{3/2}$ orbits,
which, for the SV functional, in the double magic nucleus $^{40}$Ca are almost degenerate
and can therefore strongly mix, in particular
through the time-odd fields in odd-odd $^{38}$K. To gain a better insight
into the problem, in this work we perform the
NCCI study of both nuclei, $^{38}$K and $^{38}$Ca.
For our preliminary results presented in this work,
we were able to converge three low-lying antialigned
reference configurations in $^{38}$K and four configurations in
$^{38}$Ca. Their basic properties, including labels in terms of the dominant Nilsson
components of the hole orbitals, are listed in Table~\ref{t2}.

\begin{table}[tbh]
\caption{Properties of reference Slater determinants in $^{38}$K and
$^{38}$Ca nuclei, including their excitation energies, valence particle
alignments in odd-odd nucleus $^{38}$K and their orientations,
quadrupole moments, and triaxiality. The determinants are labeled by
Nilsson quantum numbers $[N,n_z,\Lambda,\Omega]$ pertaining
to dominant components of the hole states.}
\label{t2}
\renewcommand{\arraystretch}{1.3}
\centering\begin{tabular}{cccccccccc}
\hline
$k$ & $|^{38}$ K$;k\rangle$        & $\Delta$E$_{\rm HF}$  & $j_\nu / j_\pi$&   $Q_2$   &  $\gamma$       &
      $|^{38}$Ca$;k\rangle$        & $\Delta$E$_{\rm HF}$  &  $Q_2$  &  $\gamma$      \\
    &                              &       (MeV)           &                 &  (fm$^2$)&   ($^\circ$)    &
                                   &       (MeV)           & (fm$^2$)        &  ($^\circ$) \\
\hline
 1  &  $|202\ttalf\rangle^{-2}$     &       0.000     &       -0.50/0.50(Y)       &      0.44       &    60   &
       $|200\thalf\rangle^{-2}$     &       0.000     &        0.47               &      60                            \\
 2  &  $|220\thalf\rangle^{-2}$     &       1.380     &       0.50/-0.50(Z)       &      0.18       &     0   &
       $|200\thalf\rangle^{-2}$     &       0.762     &        0.03               &       0                    \\
 3  &  $|211\thalf\rangle^{-2}$     &       1.559     &       -1.50/1.50(Z)       &      0.22       &     0   &
       $|211\thalf\rangle^{-2}$     &       1.669     &        0.24               &       0                    \\
 4  &                              &                 &                           &                 &         &
$|220\thalf\rangle^{-1}\otimes|202\ttalf\rangle^{-1}$     &       2.903     &        0.09               &      60                    \\
\hline
\end{tabular}
\end{table}

Results of our NCCI calculations, giving energies of the $0^+$ states and
the corresponding ISB corrections to $\beta$-decays,
are visualized in Fig.~\ref{f3}. Again, our model
accurately reproduces the experimental excitation energy of the second $0^+_2$ state
in $^{38}$Ca. Indeed, the measured value, $\Delta E_{\rm EXP}=3057(18)$\,keV, is only 186\,keV
higher than the calculated one, $\Delta E_{\rm TH}=2871$\,keV. The ISB
corrections to the $^{38}$Ca$\rightarrow$$^{38}$K transitions are for
$0_1^+\rightarrow 0_1^+$ and $0_2^+\rightarrow 0_2^+$ equal to 1.7\%
and 1.5\%, respectively. As compared to the static theory, which for the $0_1^+$
states gives $\delta_{\rm C}$=8.9\%, these values are strongly
reduced, but they are almost twice larger than the
result of TH~\cite{(Tow08)}, who quote $\delta_{\rm C}$=0.745(70)\%.

Let us finally mention that the calculated energies of
$0_2^+$ relative to $0_1^+$ states in $^{38}$K and $^{38}$Ar
(preliminary value resulting from mixing of $0^+$ states projected from three HF
configurations) are $\Delta E_{\rm TH}=2757$\,keV and $\Delta E_{\rm
TH}=3161$\,keV, respectively. The latter value is in very good
agreement with the experimental relative energy equal to $\Delta
E_{\rm EXP}=3377.45(12)$\,keV.

\section{Summary}\label{sec4}

We presented a novel no-core configuration-interaction approach,
which is based on mixing the isospin- and
angular-momentum-projected deformed DFT configurations. The model is
specifically tailored to determining the low-lying spectra and
$\beta$-decay transitions in $N\approx Z$ nuclei, where the isospin degree of
freedom is essential. The model can be viewed as a variant of
the no-core shell model. Its advantage over the standard
shell model formulation is that it can be applied, at least in principle, to
any nucleus of arbitrary mass and number parity.

Two applications of the model, both relevant to studies of
superallowed Fermi $\beta$ decays, were presented. The model has been
used to compute the $0^+$ spectrum of $^{62}$Zn and ISB corrections
to the Fermi $\beta$ decays between the $0_1^+ \rightarrow 0_1^+$ and
$0_2^+ \rightarrow 0_2^+$ isobaric analogue states in $^{38}$Ca and
$^{38}$K. We demonstrated that without adjusting any single
parameter, it well reproduces the spectra of $0^+$ states.
Predictions for the ISB corrections appear to be, at least at
present, somewhat less reliable. The reason is that the isospin
mixing is a very subtle effect, requiring a perfect matching of
spaces of states used in the parent and daughter nuclei, which is
difficult to achieve in practice. Work toward improving this aspect
of the model is in progress.

\vspace{0.3cm}

This work was supported in part by the Polish National Science Centre
(NCN) under Contract No. 2012/07/B/ST2/03907, by the THEXO JRA within
the EU-FP7-IA project ENSAR (No.\ 262010), by the ERANET-NuPNET grant
SARFEN of the Polish National Centre for Research and Development,
and by the Academy of Finland and University of Jyv\"askyl\"a within
the FIDIPRO programme.


\bibliographystyle{jpsj}
%
%
%


\end{document}